\documentclass[pre,twocolumn,superscriptaddress,amsfonts]{revtex4}

\usepackage{bm,amsmath,dcolumn}
\usepackage[dvips]{graphicx}

\newcommand\qq{\bm{q}}
\newcommand\rr{\bm{r}}
\newcommand\uu{\bm{u}}
\newcommand\vv{\bm{v}}
\newcommand\xx{\bm{x}}
\newcommand\yy{\bm{y}}
\newcommand\JJ{\bm{J}}
\renewcommand\SS{\bm{S}}
\newcommand\ttau{\bm{\tau}}
\newcommand\xxi{\bm{\xi}}
\newcommand\zzeta{\bm{\zeta}}

\newcommand\cG{{\mathcal{G}}}
\newcommand\cH{{\mathcal{H}}}
\newcommand\cL{{\mathcal{L}}}

\newcommand\BSC{{\textrm{BSC}}}
\newcommand\BIAWGNC{{\textrm{BIAWGNC}}}
\newcommand\BILC{{\textrm{BILC}}}

\newcommand\RS{{\textrm{RS}}}
\newcommand\ferro{{\textrm{ferro}}}
\newcommand\subopt{{\textrm{sf}}}
\newcommand\para{{\textrm{para}}}
\newcommand\C{{\textsf{C}}}

\newcommand\sign{\mathop{\mathrm{sign}}\nolimits}

\newcommand\Extr{\mathop{\mathrm{Extr}}}

\begin{document}
\title{Typical performance of low-density parity-check codes\\
  over general symmetric channels}
\author{Toshiyuki Tanaka}
\affiliation{Department of Electronics and Information Engineering, 
  Tokyo Metropolitan University, 1-1 Minami-Osawa, Hachioji-shi, Tokyo,
  192-0397 Japan}
\affiliation{Neural Computing Research Group, Aston University, 
  Aston Triangle, Birmingham, B4 7ET, United Kingdom}
\author{David Saad}
\affiliation{Neural Computing Research Group, Aston University, 
  Aston Triangle, Birmingham, B4 7ET, United Kingdom}

\date{\today}

\begin{abstract}
Typical performance of low-density parity-check (LDPC) codes 
over a general binary-input output-symmetric memoryless channel 
is investigated using methods of statistical mechanics. 
Theoretical framework for dealing with general symmetric channels 
is provided, based on which 
Gallager and MacKay-Neal codes are studied as 
examples of LDPC codes.  
It has been shown that the basic properties of these codes 
known for particular channels, 
including the property to potentially saturate 
Shannon's limit, hold for general symmetric channels. 
The binary-input additive-white-Gaussian-noise channel 
and the binary-input Laplace channel 
are considered as specific channel noise models. 
\end{abstract}


\maketitle

\section{Introduction}
We investigate the typical performance of low-density parity-check
(LDPC) codes over a general binary-input output-symmetric (BIOS)
memoryless channel. Previous statistical physics based analyses of
LDPC codes have discovered some interesting properties, including the
fact that they can, in principle, saturate the information-theoretic
upper bound (Shannon's bound defined by the channel coding
theorem~\cite{Shannon48}) with low connectivity values. Existing
statistical mechanical studies on the LDPC codes, however, have been
mostly confined to the case of binary symmetric channel (BSC), which
fits into the statistical-mechanical framework in a natural
way~\cite{KMS00,MKSV00,VSK00,VSK02}. 
Notable exceptions are the work by Montanari~\cite{montanari} that discusses 
the case of binary-input additive-white-Gaussian-noise channel (BIAWGNC) 
as well as the BSC case and the study of Sourlas
codes~\cite{Sourlas89}, a simple LDPC code, in which non-BSC
channels are addressed~\cite{Rujan93,NW99,VSK99}.  From the
statistical-mechanical point of view, LDPC codes are regarded as
random spin systems; it is therefore natural to expect that they will
exhibit some sort of universality, just as typical
statistical-mechanical systems do, so that general properties of LDPC
codes observed in the BSC case will be preserved when different
communication channels are considered. In this paper we investigate
the properties of LDPC codes in binary-input output-symmetric channels
and show that this is generally the case. In particular, we show that
the finite connectivity LDPC codes can saturate Shannon's bound for
general BIOS channel.

The paper is organized as follows: In section~\ref{sec:framework} we
introduce the general framework, notation, codes and the channels that
we will focus on. In section~\ref{sec:analysis} we will briefly
describe the calculation for the various channels, while the results
obtained will be described in section~\ref{sec:res}, followed by the
conclusions.

\section{The general framework}
\label{sec:framework}
\subsection{Symmetric channels}
\label{sec:channel}

We consider the general class of binary-input output-symmetric (BIOS)
memoryless channel. The input of the channel is binary ($\pm1$), and
the output may take any real value. The characteristics of a channel
is described by the channel transition probabilities, $P(y|x=1)$ and
$P(y|x=-1)$. Let $p(y)\equiv P(y|x=1)$. A symmetric channel is
characterized as a channel whose transition probabilities satisfy
$P(y|x=-1)=P(-y|x=1)=p(-y)$. Various types of channel models of
practical interest fall into the class of BIOS channels, including the
binary symmetric channel (BSC)
\begin{equation}
  p_{\BSC}(y)=(1-p)\delta(y-1)+p\delta(y+1),
\end{equation}
the binary-input additive-white-Gaussian-noise channel (BIAWGNC) 
\begin{equation}
  p_{\BIAWGNC}(y)=\frac{1}{\sqrt{2\pi\sigma^2}}
  e^{-{(y-1)^2/2\sigma^2}},
\end{equation}
and the binary-input Laplace channel (BILC) 
\begin{equation}
  p_{\BILC}(y)=\frac{1}{\lambda}e^{-|y-1|/\lambda},
\end{equation}
Each of the
parameters $p$, $\sigma^2$, and $\lambda$ represents the degree of
degradation induced by the channel noise. We call each
of them the noise level and let $d$ denotes the generic one.

An apparent technical difficulty in dealing with 
a general channel of real-valued output 
is that it is not at all obvious how to define 
the syndrome from the received signal: 
The modulo 2 arithmetic involved in computing 
syndrome in the BSC case is not directly applicable 
to the cases of real-valued received signal. 
This difficulty is resolved by using a truncation
procedure~\cite{MacKay99}: We conceptually consider another {\em
fictitious} binary-input binary-output channel in addition to the
channel under study. Let $r$ be the (fictitious) output symbol of
this fictitious channel. We can assign to $r$ either of the values
$\pm1$ arbitrarily, and the binary channel noise $\zeta$ for the
fictitious channel is defined therefrom, via $r=x\zeta$. For the sake
of making the argument simple, we assign $r=1$ without loss of
generality. Since the prior probability of $\zeta$ ({\em before}
receiving $y$) should be $P(\zeta=\pm1)=1/2$, the joint distribution
of $y$ and $\zeta$ is given by
\begin{equation}
  P(y,\zeta)=\frac{p(\zeta y)}{2}
\end{equation}
since the truncation procedure used here yields $x=\zeta$.

\subsection{Gallager code}
\label{sec:Gallager}

LDPC codes have been originally introduced by Gallager in his seminal
work from 1963~\cite{Gallager62}. Gallager's original
construction~\cite{Gallager62} is one of the most extensively studied
LDPC codes in the information theory literature. It is defined by its
parity-check matrix $A=[C_1|C_2]$ of dimensionality $(M-N)\times M$,
which is taken to be random and very sparse. The submatrix $C_2$, of
dimensionality $(M-N)\times(M-N)$, is assumed invertible.

In the encoding step, the encoder computes a codeword 
from the information vector $\xxi\in\{0,1\}^N$ by employing a generator 
matrix $G$
\begin{equation}
  \xx=G^T\xxi\mod 2,
\end{equation}
where the generator matrix is defined by 
\begin{equation}
  G=[I|C_2^{-1}C_1]\mod 2.
\end{equation}
This construction ensures $AG^T=0\mod 2$. 
The information code rate for unbiased messages is $R=N/M$. 

In regular Gallager codes, the number of non-zero elements per row of
$A$ is fixed to be $K$. We call it the row constraint. Average
number of non-zero elements per column is then $C\equiv K(M-N)/M$,
whereas we will consider the case in which the number of non-zero
elements in each column is forced to be exactly $C$, which we term the
{\em column constraint}. {\em Irregular} Gallager codes can be
defined by relaxing these constraints. It has been known that making
code construction irregular may improve performance
significantly~\cite{RSU01}, but we will not discuss irregular codes in
the current paper. We call the resulting regular Gallager code a
$(C,K)$-Gallager code.

\subsection{MN code}
\label{sec:MN}

We also discuss a variant of LDPC codes, called the MN
code~\cite{MN95,MacKay99}. The generator matrix $G^T$ of the MN code
is defined by
\begin{equation}
  G^T=C_n^{-1}C_s\mod 2,
\end{equation}
where $C_s$ and $C_n$ are sparse matrices of dimensionality $M\times
N$ and $M\times M$, respectively; $C_n$ is assumed invertible. The
information rate for the code is $R=N/M$ for unbiased message.

In regular MN codes the row and column constraints are imposed on both
matrices $C_s$ and $C_n$. The number of non-zero elements per row
of $C_s$ and $C_n$ should be exactly $K$ and $L$, respectively.
Also here, we do not discuss irregular MN codes~\cite{KS99} in this
paper. The number of non-zero elements per column of $C_s$ and
$C_n$ are set to $C$ and $L$, respectively, where $C=KM/N$ holds. We
call the resulting code a $(K,C,L)$-MN code.

\section{Analysis}
\label{sec:analysis}

\subsection{Gallager code}
\label{sec:rep-gal}

The basic idea behind the statistical-mechanical treatment of the LDPC
codes is the equivalence between the decoding problem and the thermal
equilibrium distribution of a dilute Ising spin system. In order to
see this in the Gallager code case, one should first note that the
decoding problem is to find $\ttau$ which is best supported (i.e.,
most probable) by the received signal $\yy$ among the set of $\ttau$
satisfying the parity-check equation ($A\zzeta=A\ttau\mod 2$ if we
write it in the $\{0,1\}$-notation). The set is expressed as
\begin{equation}
  \biggl\{\ttau\biggm|\lim_{\gamma\to\infty}
  \exp\biggl[-\gamma\sum_{\mu=1}^{M-N}
  \biggl(J_\mu\prod_{j\in\cL(\mu)}\tau_j-1\biggr)\biggr]=1
  \biggr\},
\end{equation}
where 
\begin{equation}
  \cL(\mu)=\{j|A_{\mu j}=1\}
\end{equation}
denotes the set of indices for which 
the parity-check matrix $A$ has 1's in $\mu$-th row, and 
\begin{equation}
  J_\mu=\prod_{j\in\cL(\mu)}\zeta_j
\end{equation}
is $\mu$-th check. The posterior probability of $\ttau$ conditioned
on the received signal $\yy$ then acquires the following
Gibbs-Boltzmann form:
\begin{equation}
  P_\gamma(\ttau|\yy)=\frac{1}{Z}
  \exp\bigl[-\beta\cH_\gamma(\ttau;\yy,\JJ)\bigr]
\end{equation}
in which we have to take the limit $\gamma\to\infty$ 
and consider it at $\beta=1$ (Nishimori's 
temperature~\cite{Rujan93,Nishimori93,Iba99,nishimori_book}) 
in order to obtain the true posterior. 
The Hamiltonian $\cH_\gamma(\ttau;\yy,\JJ)$ is 
defined as 
\begin{eqnarray}
  \cH_\gamma(\ttau;\yy,\JJ)
  &=&
  -\gamma\sum_{\mu=1}^{M-N}
  \biggl(J_\mu\prod_{j\in\cL(\mu)}\tau_j-1\biggr)
  \nonumber\\
  &&{}-\sum_{j=1}^M\log p(\tau_j y_j),
\end{eqnarray}
The channel characteristics enters into the Hamiltonian 
as the term $\log p(\tau_jy_j)$ which, 
by noting that $\tau_j$ takes $\pm1$, 
can be rewritten as 
\begin{equation}
  \log p(\tau_jy_j)=
  \tau_j\frac{1}{2}\log\frac{p(y_j)}{p(-y_j)}
  +\frac{1}{2}\log p(y_j)p(-y_j).
\end{equation}
From this expression it immediately follows that it is the
log-likelihood ratio $h_j\equiv(1/2)\log(p(y_j)/p(-y_j))$ of the
channel noise $y_j$ which serves as the external field acting on site
$j$, and that the channel characteristics defines the field
distribution. Analyzing the effect of having different communication
channels on the code properties, therefore reduces to
investigating the effect of different field distributions on the physical
properties of the system. The field distributions $p(h)$ for various
channel models are as follows:
\begin{itemize}
\item BSC:
  \begin{eqnarray}
    p_{\BSC}(h)&=&(1-p)\delta\biggl(h-\frac{1}{2}\log\frac{1-p}{p}\biggr)
    \nonumber\\
    &&{}+p\delta\biggl(h+\frac{1}{2}\log\frac{1-p}{p}\biggr)
  \end{eqnarray}
\item BIAWGNC:
  \begin{equation}
    p_{\BIAWGNC}(h)=\sqrt{\frac{\sigma^2}{2\pi}}
      e^{-(h-\sigma^{-2})^2/2\sigma^{-2}}
  \end{equation}
\item BILC:
  \begin{eqnarray}
    p_{\BILC}(h)&=&\frac{1}{2}\delta(h-\lambda^{-1})
    +\frac{e^{-2\lambda^{-1}}}{2}\delta(h+\lambda^{-1})
    \nonumber\\
    &&{}+\chi[-\lambda^{-1}<h<\lambda^{-1}]
    \frac{1}{2}e^{h-\lambda^{-1}},
  \end{eqnarray}
  where $\chi[X]$ is the indicator function, 
  taking 1 when $X$ is true and 0 otherwise. 
\end{itemize}
\begin{figure}
  \centering
  \begin{minipage}{45mm}
    \centering
    \includegraphics[width=45mm]{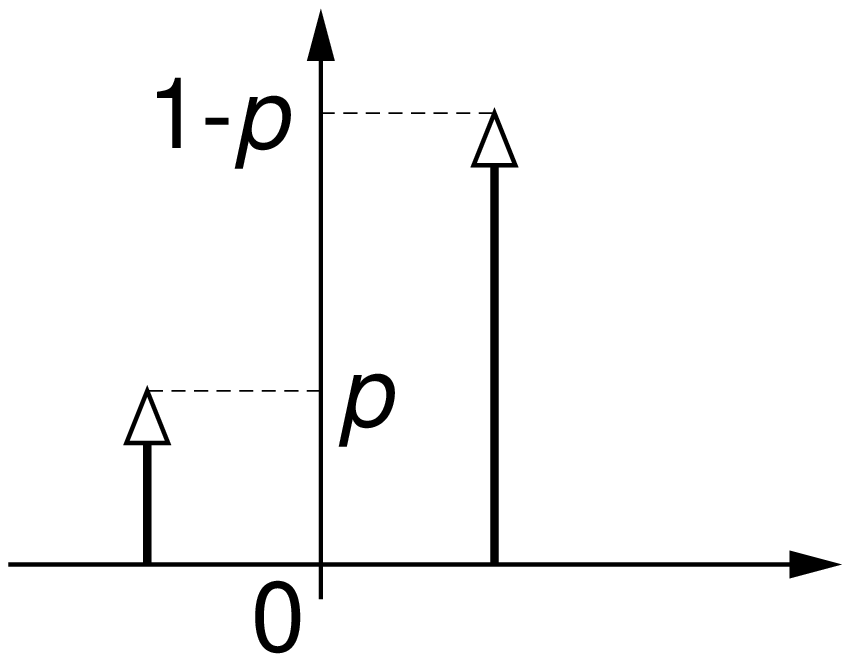}\\
    (a)\ BSC
  \end{minipage}\ %
  \begin{minipage}{45mm}
    \centering
    \includegraphics[width=45mm]{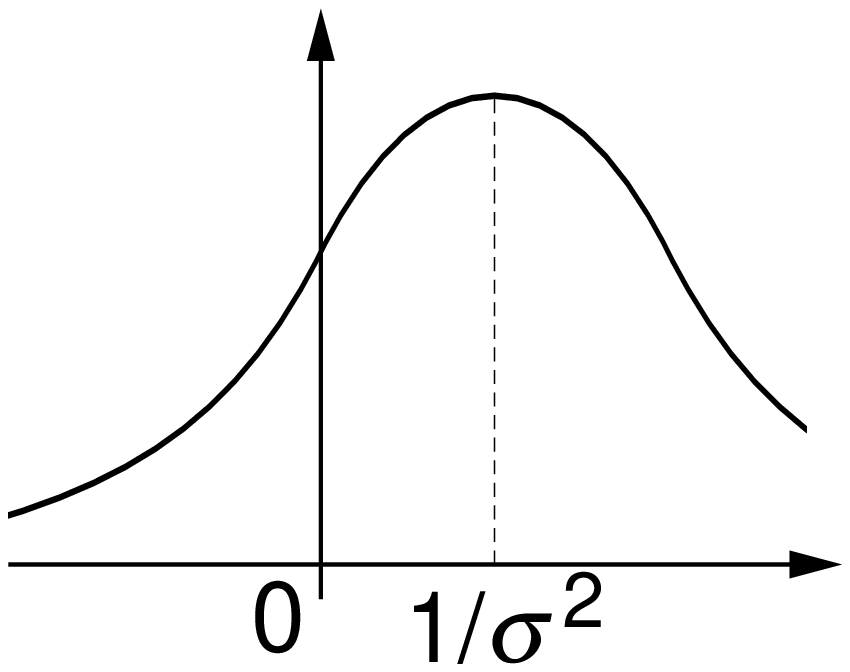}\\
    (b)\ BIAWGNC
  \end{minipage}\ %
  \begin{minipage}{45mm}
    \centering
    \includegraphics[width=45mm]{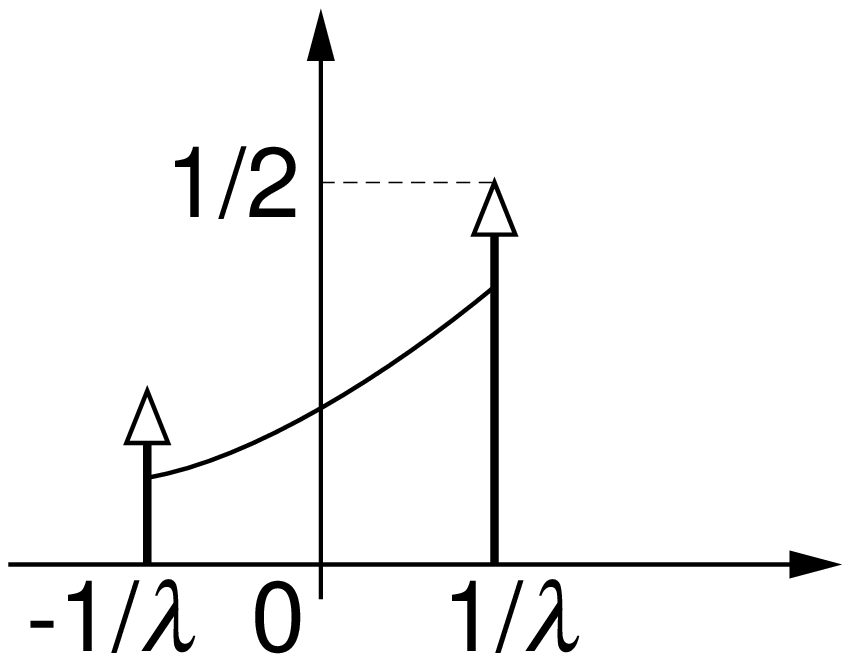}\\
    (c)\ BILC
  \end{minipage}\ %
  \caption{Field distributions corresponding to various BIOS channels.}
  \label{fig:fd}
\end{figure}
Sketches of these field distributions are given in Fig.~\ref{fig:fd}. 

We assume that the free energy of the system is self-averaging, that
is,
\begin{equation}
  f=-\frac{1}{\beta}\lim_{M\to\infty}M^{-1}
  \langle\log Z\rangle_{A,\yy},
\end{equation}
and evaluate the average $\langle\cdot\rangle_{A,\yy}$ over the
received signal $\yy$ and the randomness of the parity-check matrix
$A$ using the replica method,
\begin{equation}
  f=-\frac{1}{\beta}\lim_{M\to\infty}
  \lim_{n\to0}M^{-1}
  \frac{\partial}{\partial n}\log\langle Z^n\rangle_{A,\yy}.
\end{equation}
In calculating the free energy, we perform the gauge transformation
$\tau_j\to\zeta_j\tau_j$, $y_j\to\zeta_j y_j$.  The average over $\yy$
can be taken with respect to $\prod_{j=1}^Mp(y_j)$ after having
performed the gauge transformation.  We have to introduce a random
tensor to take average over $A$.

Following basically the same procedure as in~\cite{MKSV00} and
exchanging the order of the two limits, taking the limit $M\to\infty$
first, one obtains
\begin{equation}
  f=-\frac{1}{\beta}\lim_{n\to0}\frac{\partial}{\partial n}
  \Extr_{\qq,\hat\qq}\biggl[
    \frac{C}{K}\cG_1(\qq)-\cG_2(\qq,\hat\qq)+\cG_3(\hat\qq)
  \biggr],
\end{equation}
where 
\begin{eqnarray}
  \cG_1(\qq)&\equiv&\log
  \sum_{m=0}^n\sum_{\langle\alpha_1\cdots\alpha_m\rangle}
  q_{\alpha_1\cdots\alpha_m}^K-n\log2,
  \nonumber\\
  \cG_2(\qq,\hat\qq)&\equiv&
  \sum_{m=0}^n\sum_{\langle\alpha_1\cdots\alpha_m\rangle}
  q_{\alpha_1\cdots\alpha_m}
  \hat q_{\alpha_1\cdots\alpha_m},
  \nonumber\\
  \cG_3(\hat\qq)&\equiv&
  \log\Biggl[
    \sum_{\tau^1,\ldots,\tau^n}
    \left\langle\prod_{\alpha=1}^np(\tau^\alpha y)\right\rangle_y
    \nonumber\\
    &&
    \times\frac{1}{C!}
    \biggl(
      \sum_{m=0}^n\sum_{\langle\alpha_1\cdots\alpha_m\rangle}
      \hat q_{\alpha_1\cdots\alpha_m}
      \tau^{\alpha_1}\cdots\tau^{\alpha_m}
    \biggr)^C
  \Biggr].
  \nonumber\\
\end{eqnarray}

To proceed further we adopt the replica-symmetric (RS) ansatz and let
\begin{equation}
  \label{eq:RS}
  q_{\alpha_1\cdots\alpha_m}=q_0\int u^m\pi(u)\,du,
  \quad
  \hat q_{\alpha_1\cdots\alpha_m}=\hat q_0\int\hat u^m\hat\pi(\hat u)\,d\hat u.
\end{equation}
We will use the following simplifying notation. 
\begin{equation}
  \pi^K(\uu)\,d\uu\equiv\prod_{j=1}^K\pi(u_j)\,du_j
\end{equation}
The replica-symmetric free energy $f^\RS$ becomes 
\begin{eqnarray}
  f^\RS&=&\frac{1}{\beta}\Extr_{\pi,\hat\pi}
  \Biggl\{
  \frac{C}{K}\log 2
  \nonumber\\
  &+&C\iint\log(1+u\hat u)\,\pi(u)\,\hat\pi(\hat u)\,du\,d\hat u
  \nonumber\\
  &-&
  \frac{C}{K}\int
  \log\Bigl(1+\prod_{j=1}^Ku_j\Bigr)
  \,\pi^K(\uu)\,d\uu
  \nonumber\\
  &-&\int
  \biggl\langle
    \log\biggl[p(y)\prod_{l=1}^C(1+\hat u_l)
    +p(-y)\prod_{l=1}^C(1-\hat u_l)\biggr]
  \biggr\rangle_y
  \nonumber\\
  &&\hphantom{\int}\times
  \hat\pi^C(\hat\uu)\,d\hat\uu
  \Biggr\},
\end{eqnarray}
in which $q_0$ and $\hat q_0$ have been eliminated 
using the extremization condition $q_0\hat q_0=C$. 
Heuristic construction of a sufficient condition 
to the extremization problem 
with respect to $\pi$ and $\hat\pi$ is possible, 
and it gives the following saddle-point equations. 
\begin{eqnarray}
  \label{eq:sp-gal}
  \pi(u)
  &=&\int
  \biggl\langle\delta\biggl[
    u-\tanh\biggl(h(y)
    +\sum_{l=1}^{C-1}\tanh^{-1}\hat u_l\biggr)
  \biggr]\biggr\rangle_y
  \nonumber\\
  &&\hphantom{int}\times
  \hat\pi^{C-1}(\hat\uu)\,d\hat\uu
  \nonumber\\
  \hat\pi(\hat u)
  &=&\int
  \delta\biggl(\hat u-\prod_{j=1}^{K-1}u_j\biggr)
  \,\pi^{K-1}(\uu)\,d\uu
\end{eqnarray}
We have let 
\begin{equation}
  h(y)\equiv\frac{1}{2}\log\frac{p(y)}{p(-y)}.
\end{equation}
The performance of the code is quantified by 
the overlap $m=M^{-1}\sum_{k=1}^M\zeta_j\langle\tau_j\rangle$, 
which is given as 
\begin{equation}
  m=\int \sign(z)\,P(z)\,dz,
\end{equation}
where 
\begin{eqnarray}
  P(z)
  &=&\int
  \biggl\langle\delta\biggl[
    z-\tanh\biggl(h(y)
    +\sum_{l=1}^C\tanh^{-1}\hat u_l\biggr)
  \biggr]\biggr\rangle_y
  \nonumber\\
  &&\hphantom{\int}\times
  \hat\pi^C(\hat\uu)\,d\hat\uu.
\end{eqnarray}

\subsection{MN code}
\label{sec:rep-MN}

The decoding problem for the MN code is 
to find $\SS$ and $\ttau$ which are 
the best suitable in view of the received signal $\yy$ 
among the sets of $\SS$ and $\ttau$ 
satisfying the parity-check equation 
($C_s\SS+C_n\ttau=C_s\xxi+C_n\zzeta\mod 2$ 
if written in the $\{0,1\}$-notation). 
Defining the $\mu$th component of the check $\JJ$ as 
\begin{equation}
  J_\mu=\prod_{j\in\cL_s(\mu)}\xi_j
        \prod_{l\in\cL_n(\mu)}\zeta_l,
\end{equation}
where 
\begin{equation}
  \cL_s(\mu)=\{j|(C_s)_{\mu j}=1\},
  \quad
  \cL_n(\mu)=\{l|(C_n)_{\mu l}=1\},
\end{equation}
the posterior probability of $\SS$ and $\ttau$ 
conditioned on the received signal $\yy$ 
and the check $\JJ$ is given by 
\begin{equation}
  P_\gamma(\SS,\ttau|\yy,\JJ)
  =\frac{1}{Z}\exp\bigl[-\beta\cH_\gamma(\SS,\ttau;\yy,\JJ)\bigr],
\end{equation}
in the limit $\gamma\to\infty$ and at $\beta=1$, 
where the Hamiltonian $\cH_\gamma(\SS,\ttau;\yy,\JJ)$ 
is defined as 
\begin{eqnarray}
  \label{eq:Hamiltonian-MN}
  \cH_\gamma(\SS,\ttau;\yy,\JJ)
  &=&-\gamma\sum_{\mu=1}^M\biggl(J_\mu
  \prod_{j\in\cL_s(\mu)}S_j
  \prod_{l\in\cL_n(\mu)}\tau_l
  -1\biggr)
  \nonumber\\
  &&{}-F_s\sum_{j=1}^NS_j-\sum_{l=1}^M\log p(\tau_l y_l),
\end{eqnarray}
where $F_s$ is a parameter representing the bias of the information
vector $\xxi$ in such a way that $P(\xi_j=\pm1)=(1\pm\tanh F_s)/2$
holds. The form of Eq.~(\ref{eq:Hamiltonian-MN}) clearly shows that
the channel characteristics again acts as random field on
$\{\tau_l\}$, where the log likelihood ratio gives the actual value of
the field.

The replica calculation can be done 
along the same way as in the case of the Gallager code. 
We have performed the gauge transformation 
$S_j\to\xi_j S_j$, $\tau_j\to\zeta_j\tau_j$, 
and $y_j\to\zeta_j\tau_j$. 
The free energy $f$ becomes 
\begin{eqnarray}
  \label{eq:MN-free-energy}
  f&=&-\frac{1}{\beta}\lim_{n\to0}\frac{\partial}{\partial n}
  \Extr_{\qq,\hat\qq,\rr,\hat\rr}\biggl[
    \frac{C}{K}\cG_1(\qq,\rr)
  \nonumber\\
  &&\hphantom{-\frac{1}{\beta}\lim_{n\to0}\frac{\partial}{\partial n}
  \Extr_{\qq,\hat\qq,\rr,\hat\rr}\biggl[}
    -\cG_2(\qq,\hat\qq,\rr,\hat\rr)+\cG_3(\hat\qq,\hat\rr)
  \biggr],
  \nonumber\\
\end{eqnarray}
where 
\begin{eqnarray}
  \cG_1(\qq,\rr)&\equiv&\log
  \sum_{m=0}^n\sum_{\langle\alpha_1\cdots\alpha_m\rangle}
  q_{\alpha_1\cdots\alpha_m}^K
  r_{\alpha_1\cdots\alpha_m}^L
  \nonumber\\
  &&{}-n\log2,
  \nonumber\\
  \cG_2(\qq,\hat\qq,\rr,\hat\rr)&\equiv&
  \sum_{m=0}^n\sum_{\langle\alpha_1\cdots\alpha_m\rangle}
  q_{\alpha_1\cdots\alpha_m}
  \hat q_{\alpha_1\cdots\alpha_m}
  \nonumber\\
  &&{}+\frac{M}{N}\sum_{m=0}^n\sum_{\langle\alpha_1\cdots\alpha_m\rangle}
  r_{\alpha_1\cdots\alpha_m}
  \hat r_{\alpha_1\cdots\alpha_m},
  \nonumber\\
\end{eqnarray}
and 
\begin{eqnarray}
  \cG_3(\hat\qq,\hat\rr)
  &\equiv&
  \log\Biggl[
    \sum_{S^1,\ldots,S^n}
    \Bigl\langle e^{F_s\sum_{\alpha=1}^n\xi S^\alpha}
    \Bigr\rangle_\xi
    \nonumber\\
    &&{}\times\frac{1}{C!}
    \biggl(
      \sum_{m=0}^n\sum_{\langle\alpha_1\cdots\alpha_m\rangle}
      \!\!
      \hat q_{\alpha_1\cdots\alpha_m}
      S^{\alpha_1}\cdots S^{\alpha_m}
    \biggr)^C
  \Biggr]
  \nonumber\\
  &+&\frac{M}{N}
  \log\Biggl[
    \sum_{\tau^1,\ldots,\tau^n}
    \biggl\langle\prod_{\alpha=1}^np(\tau^\alpha y)\biggr\rangle_y
    \nonumber\\
    &&{}\times\frac{1}{L!}
    \biggl(
      \sum_{m=0}^n\sum_{\langle\alpha_1\cdots\alpha_m\rangle}
      \!\!
      \hat r_{\alpha_1\cdots\alpha_m}
      \tau^{\alpha_1}\cdots\tau^{\alpha_m}
    \biggr)^L
  \Biggr].
  \nonumber\\
\end{eqnarray}

We adopt the RS ansatz as before, under which we have 
\begin{equation}
  \label{eq:RS-MN}
  r_{\alpha_1\cdots\alpha_m}=r_0\int v^m\rho(v)\,dv,
  \quad
  \hat r_{\alpha_1\cdots\alpha_m}=\hat r_0\int\hat v^m
  \hat\rho(\hat v)\,d\hat v,
\end{equation}
in addition to Eq.~(\ref{eq:RS}). 
The replica-symmetric free energy $f^\RS$ becomes 
\begin{eqnarray}
  f^\RS&=&\frac{1}{\beta}\Extr_{\pi,\hat\pi,\rho,\hat\rho}\Biggl\{
  \frac{C}{K}\log2
  \nonumber\\
  &+&C\iint\log(1+u\hat u)\,\pi(u)\,\hat\pi(\hat u)\,du\,d\hat u
  \nonumber\\
  &+&\frac{CL}{K}\iint\log(1+v\hat v)\,\rho(v)\,\hat\rho(\hat v)\,dv\,d\hat v
  \nonumber\\
  &-&\frac{C}{K}\iint
  \log\biggl(1+\prod_{k=1}^Ku_k\prod_{l=1}^Lv_l\biggr)
  \nonumber\\
  &&\hphantom{\frac{C}{K}\iint}
  \times
  \pi^K(\uu)\,d\uu
  \,\rho^L(\vv)\,d\vv
  \nonumber\\
  &-&\int
  \biggl\langle
    \log\biggl[\sum_{S=\pm1}e^{F_s\xi S}
      \prod_{k=1}^C(1+S\hat u_k)\biggr]
  \biggr\rangle_\xi
  \hat\pi^C(\hat\uu)\,d\hat\uu
  \nonumber\\
  &-&\frac{C}{K}\int
  \biggl\langle
    \log\biggl[\sum_{\tau=\pm1}p(\tau y)
      \prod_{l=1}^L(1+\tau\hat v_l)\biggr]
  \biggr\rangle_y
  \nonumber\\
  &&\hphantom{\frac{C}{K}\int}\times
  \hat\rho^L(\hat\vv)\,d\hat\vv
  \Biggr\},
\end{eqnarray}
in which $q_0$, $\hat q_0$, $r_0$, and $\hat r_0$ have been 
eliminated using the extremization conditions, 
$q_0\hat q_0=C$ and $r_0\hat r_0=L$. 

Construction of a heuristic solution 
to the extremization problem can be done in the same manner, 
which yields the following saddle-point equations: 
\begin{eqnarray}
  \pi(u)&=&\int
  \biggl\langle\delta\biggl[
    u-\tanh\biggl(F_s\xi +\sum_{l=1}^{C-1}\tanh^{-1}\hat u_l\biggr)
  \biggr]\biggr\rangle_\xi
  \nonumber\\
  &&\hphantom{\int}\times
  \hat\pi^{C-1}(\hat\uu)\,d\uu
  \nonumber\\
  \hat\pi(\hat u)&=&\iint
  \delta\biggl(
    \hat u-\prod_{k=1}^{K-1}u_k\prod_{l=1}^Lv_l
  \biggr)
  \,\pi^{K-1}(\uu)\,d\uu
  \,\rho^L(\vv)\,d\vv
  \nonumber\\
  \rho(v)&=&\int
  \biggl\langle\delta\biggl[
    v-\tanh\biggl(h(y)
    +\sum_{l=1}^{L-1}\tanh^{-1}\hat v_l\biggr)
  \biggr]\biggr\rangle_y
  \nonumber\\
  &&\hphantom{\int}\times
  \hat\rho^{L-1}(\hat\vv)\,d\hat\vv
  \nonumber\\
  \hat\rho(\hat v)&=&\iint
  \delta\biggl(\hat v-\prod_{k=1}^Ku_k\prod_{l=1}^{L-1}v_l\biggr)
  \,\pi^K(\uu)\,d\uu
  \,\rho^{L-1}(\vv)\,d\vv
  \nonumber\\
\end{eqnarray}

The overlap is then evaluated by 
\begin{equation}
  m=\int \sign(z)\,P(z)\,dz,
\end{equation}
where 
\begin{eqnarray}
  P(z)&=&\int
  \biggl\langle\delta\biggl[
    z-\tanh\biggl(F_s\xi +\sum_{l=1}^C\tanh^{-1}\hat u_l\biggr)
  \biggr]\biggr\rangle_\xi
  \nonumber\\
  &&\hphantom{\int}\times
  \hat\pi^C(\hat\uu)\,d\hat\uu.
\end{eqnarray}

It is worthwhile mentioning that, when the message is unbiased
($F_s=0$) and $K$ is even, saddle-point solutions have the following
symmetry: For each solution $\{\pi(u),\hat\pi(\hat
u),\rho(v),\hat\rho(\hat v)\}$ there is another solution
$\{\pi(-u),\hat\pi(-\hat u),\rho(v),\hat\rho(\hat v)\}$.  The latter
has the same overlap as that of the former with the opposite sign.

\section{Results}
\label{sec:res}

\subsection{Gallager code}
\label{sec:res-gal}

\subsubsection{Analytical solutions}

Of particular interest is the ferromagnetic state, 
which corresponds to an error-free communication. 
One can see that the assertion 
\begin{equation}
  \pi(u)=\delta(u-1),\quad\hat\pi(\hat u)=\delta(\hat u-1)
\end{equation}
always satisfies the saddle-point equation~(\ref{eq:sp-gal}) 
irrespective of the values of $K$ and $C$ 
(provided that $K,C\ge2$), 
thereby providing a solution. 
The overlap and the free energy of the solution 
at $\beta=1$ are $m_\ferro=1$ 
and $f_\ferro=-\langle\log p(y)\rangle_y$, respectively. 
One can therefore identify this as the ferromagnetic solution. 

Another solution, which can be found in the limit $K\to\infty$, is the
sub-optimal ferromagnetic solution
\begin{equation}
  \pi(u)=\left\langle\delta\bigl[
    u-\tanh h(y)
  \bigl]\right\rangle_y,
  \quad\hat\pi(\hat u)=\delta(\hat u),
\end{equation}
for which 
\begin{equation}
  m_\subopt=\bigl\langle\sign\bigl[p(y)-p(-y)\bigr]\bigr\rangle_y
\end{equation}
and 
\begin{equation}
  f_\subopt=\frac{C}{K}\log2
  -\bigl\langle\log\bigl[p(y)+p(-y)\bigr]\bigr\rangle_y.
\end{equation}
The difference of the free energy is expressed as 
\begin{equation}
  f_\subopt-f_\ferro=\C-R\log2,
\end{equation}
where $\C$ is the channel capacity of the BIOS channel defined as 
\begin{equation}
  \C=\log 2-\bigl\langle\log\bigl[p(y)+p(-y)\bigr]\bigr\rangle_y
  +\bigl\langle\log p(y)\bigr\rangle_y.
\end{equation}
This proves that the thermodynamic transition between the
ferromagnetic and sub-optimal ferromagnetic solutions (no other
solution has been identified in this case) occurs at the theoretical
limit.  This means that the maximum rate $R_{\rm max}$, up to which
error-free communication is theoretically possible, asymptotically
achieves the theoretical limit as $K\to\infty$.  This result has been
known for BSC channel~\cite{VSK00,VSK02} in the physics literature and
is in agreement with results reported in the information theory
literature~\cite{MacKay99}. The current result is an extension to the
case of a general BIOS channel.

\subsubsection{Numerical solutions of saddle-point equations}

\begin{figure}
  \centering
  \includegraphics[width=80mm]{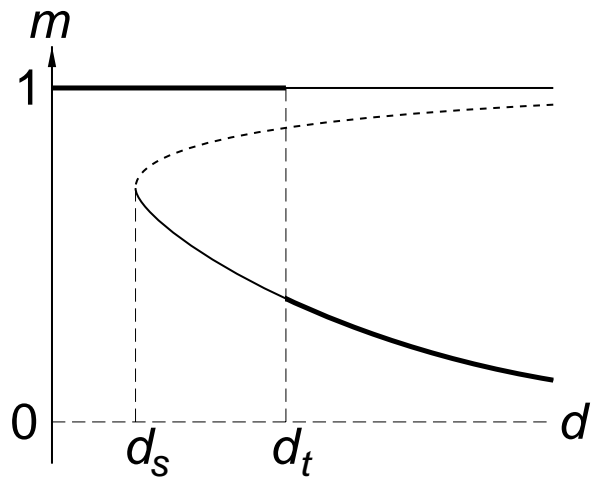}
  \caption{Noise-overlap diagram for Gallager code.  Thick solid
  lines stand for the stable state, thin solid lines for metastable
  state, and broken lines for unstable states. The ferromagnetic
  solution is characterized by the $m=1$ solution, while $m<1$
  defines the suboptimal ferromagnetic solution.}  \label{fig:pd-gal}
\end{figure}

In finite-$K$ cases no simple analytical solution exists 
other than the ferromagnetic one, so one has
to solve the saddle-point equations numerically.  We have done it for
BIAWGNC and BILC.  The dependence of the overlap $m$ on the noise
level $d$ ($\sigma^2$ for BIAWGNC, and $\lambda$ for BILC) is
qualitatively the same as that observed in BSC: For $K\ge3$ the
ferromagnetic solution is locally stable over the whole range of noise
levels.  At $d=d_s$, another solution with $m<1$ appears, which
defines the spinodal point.  At a higher noise level $d=d_t>d_s$
thermodynamic transition takes place, beyond which the ferromagnetic
solution with $m=1$ becomes metastable (see Fig.~\ref{fig:pd-gal}).
Table~\ref{tab:Gal-BIAWGNC} summarizes the results for the BIAWGNC
case, showing the spinodal point $\sigma_s^2$ (the value of
the variance at which new, non ferromagnetic, solutions emerge), the
thermodynamic transition point $\sigma_t^2$ (at which the
thermodynamic transition occurs), and $\sigma_0^2$, the
information-theoretic upper bound of the variance allowing error-free
communication.

Table~\ref{tab:Gal-BILC} summarizes the results for the BILC case,
showing the values of
the spinodal point $\lambda_s$, the thermodynamic transition point
$\lambda_t$, and the information-theoretic upper bound $\lambda_0$.

It should be noted that 
the results for the spinodal point 
agree well with the results 
obtained by the density evolution approach~\cite{RU01}, 
as expected, since the saddle-point equations 
by the replica analysis 
happen to coincide with the time evolution equations 
in the density evolution. 

\begin{table}
  \caption{The variances $\sigma_s^2$ and $\sigma_t^2$ 
    at the spinodal point and thermodynamic transition, respectively,  
    for the BIAWGNC for various code parameters; 
    $\sigma_0^2$, denoting the information-theoretical upper bound 
    for error-free communication, is also shown.}
  \label{tab:Gal-BIAWGNC}
  \begin{ruledtabular}
  \begin{tabular}{rrdddd}
    $C$ & $K$ & \mbox{$R$} 
    & \mbox{$\sigma_s^2$} & \mbox{$\sigma_t^2$} 
    & \mbox{$\sigma_0^2$} \\\hline
     3 &  6 & 0.5   & 0.775 & 0.899 & 0.958 \\
     4 &  8 & 0.5   & 0.701 & 0.943 & 0.958 \\
     5 & 10 & 0.5   & 0.629 & 0.952 & 0.958 \\
     3 &  5 & 0.4   & 1.017 & 1.253 & 1.321 \\
     4 &  6 & 0.333 & 1.020 & 1.666 & 1.681 \\
     3 &  4 & 0.25  & 1.598 & 2.325 & 2.401
  \end{tabular}
  \end{ruledtabular}
\end{table}

\begin{table}
  \caption{The parameter values $\lambda_s$ and $\lambda_t$ 
    at the spinodal point and thermodynamic transition, respectively,  
    for the BILC with various code parameters;
    $\lambda_0$, denoting the information-theoretical upper bound 
    for error-free communication, is also shown.}
  \label{tab:Gal-BILC}
  \begin{ruledtabular}
  \begin{tabular}{rrdddd}
    $C$ & $K$ & \mbox{$R$} 
    & \mbox{$\lambda_s$} & \mbox{$\lambda_t$} 
    & \mbox{$\lambda_0$} \\\hline
     3 &  6 & 0.5   & 0.651 & 0.712 & 0.752 \\
     4 &  8 & 0.5   & 0.618 & 0.741 & 0.752 \\
     5 & 10 & 0.5   & 0.581 & 0.746 & 0.752 \\
     3 &  5 & 0.4   & 0.773 & 0.875 & 0.914 \\
     4 &  6 & 0.333 & 0.782 & 1.045 & 1.055 \\
     3 &  4 & 0.25  & 1.018 & 1.260 & 1.298
  \end{tabular}
  \end{ruledtabular}
\end{table}


\subsection{MN code}
\label{sec:res-MN}

\subsubsection{Analytical solutions}

In the following we restrict our discussion of the MN code to the
unbiased case $F_s=0$.  The ferromagnetic solution, corresponding to
the error-free communication, can be constructed for the MN code with
$L\ge2$.  (In fact, in the case $L=1$ the matrix $C_n$ reduces to a
simple permutation matrix, so that we have to estimate each element of
noise separately.  This case is not at all interesting and therefore
we will not discuss it any more.)  It is given by
\begin{eqnarray}
  \label{eq:MN-para-sol}
  &&\pi(u)=\delta(u-1),\quad
  \hat\pi(\hat u)=\delta(\hat u-1),
  \nonumber\\
  &&\rho(v)=\delta(v-1),\quad
  \hat\rho(\hat v)=\delta(\hat v-1),
\end{eqnarray}
for which $m_\ferro=1$ and 
\begin{equation}
  f_\ferro=-\frac{C}{K}\bigl\langle\log p(y)\bigr\rangle_y. 
\end{equation}
The MN code has the following paramagnetic solution 
for $K\ge2$: 
\begin{equation}
  \begin{array}{ll}
  \pi(u)=\delta(u),
  &\hat\pi(\hat u)=\delta(\hat u),
  \\
  \rho(v)=\left\langle\delta\bigl[
    v-\tanh h(y)
  \bigl]\right\rangle_y,\quad{}
  &\hat\rho(\hat v)=\delta(\hat v),
  \end{array}
\end{equation}
which yields $m_\para=0$ and 
\begin{equation}
  f_\para=\biggl(\frac{C}{K}-1\biggr)\log2
  -\frac{C}{K}\bigl\langle\log\bigl[p(y)+p(-y)\bigr]\bigr\rangle_y.
\end{equation}
Again, since 
\begin{equation}
  f_\para-f_\ferro=\frac{C}{K}(\C-R\log2)
\end{equation}
holds, we conclude that for the MN code the maximum rate $R_{\rm
 max}$, theoretically allowing error-free communication, achieves
the theoretical limit as long as $K\ge2$, $L\ge2$, provided that
there is no locally stable solution other than the ferromagnetic and
paramagnetic solutions.  This result is an extension of the result
reported in~\cite{KMS00,MKSV00} to the case of a general BIOS
channel.

It should be noted that 
the paramagnetic solution~(\ref{eq:MN-para-sol}) is also valid 
in the limit $L\to\infty$ for the case $K=1$. 
This means that the above-mentioned result also holds 
for the case of $K=1$ asymptotically in the limit $L\to\infty$.

\subsubsection{Numerical solutions of saddle-point equations}

In order to explore solutions other than the ferromagnetic 
and paramagnetic solutions, 
we have to solve the saddle-point equations numerically. 
We have done it for the BIAWGNC and BILC cases. 
We observed qualitatively the same characteristics 
as those reported in~\cite{MKSV00}. 

The obtained numerical results suggest that the qualitative physical
properties are categorized into three types according to the $K$
value: cases with $K=1$, $K=2$ and $K\ge3$, whereas it is only
affected quantitatively by the values of $C$ and $L$, as described in
the following.

The structure of noise-overlap diagram for the MN code with $K=1$ is
qualitatively the same as that for Gallager code (see
Fig.~\ref{fig:pd-gal}): At very low noise level only the ferromagnetic
solution with $m=1$ exists.  At a certain noise level $d=d_s$ another
metastable solution with $m<1$ appears, and it becomes dominant beyond
$d=d_t>d_s$.  Since the latter solution is obtained only numerically,
there is no guarantee that the thermodynamical transition $d_t$ is
equal to the information-theoretical limit $d_0$.  Numerical results
show that in general $d_t$ {\em is} smaller than $d_0$: However, it is
also observed that, for fixed $C$, increasing $L$ makes $d_s$ smaller
and $d_t$ larger, the latter of which approaches the
information-theoretical limit $d_0$ as $L\to\infty$, 
as discussed at the end of the previous subsection.  Even for finite
$L$ the value of $d_t$ may be numerically very close to $d_0$,
especially when the rate $R$ is small.  These properties have already
been reported for the BSC case~\cite{MKSV00}, so that our finding
implies that they also hold for the BIAWGNC and BILC cases, revealing
some sort of universality.

\begin{table}
  \caption{The variances $\sigma_s^2$, $\sigma_t^2$, and $\sigma_b^2$ 
    at the spinodal point and thermodynamic transition, 
    and at bifurcation of paramagnetic solution, respectively,  
    for $(K,C,L)$-MN codes over the BIAWGNC and various code parameters;
    $\sigma_0^2$, denoting the information-theoretical upper bound 
    for error-free communication, is also shown.}
  \label{tab:MN-BIAWGNC}
  \begin{ruledtabular}
  \begin{tabular}{rrrddddd}
    $K$ & $C$ & $L$ & \mbox{$R$} 
    & \mbox{$\sigma_s^2$} & \mbox{$\sigma_t^2$} 
    & \mbox{$\sigma_b^2$} & \mbox{$\sigma_0^2$} \\\hline
      1 &   2 &   3 & 0.5   & 0.775 & 0.901 & -     & 0.958 \\
      1 &   2 &   4 & 0.5   & 0.703 & 0.944 & -     & 0.958 \\
      1 &   2 &   5 & 0.5   & 0.630 & 0.955 & -     & 0.958 \\
      1 &   3 &   2 & 0.333 & 1.338 & 1.423 & -     & 1.681 \\
      1 &   3 &   3 & 0.333 & 1.129 & 1.659 & -     & 1.681 \\
      1 &   3 &   4 & 0.333 & 0.913 & 1.672 & -     & 1.681 \\
      \noalign{\vspace{4pt}}
      2 &   3 &   2 & 0.667 & 0.536 & 0.587 & 0.612 & 0.588 \\
      2 &   3 &   3 & 0.667 & 0.430 & 0.588 & 0.459 & 0.588 \\
      2 &   3 &   4 & 0.667 & 0.368 & 0.588 & 0.385 & 0.588 \\
      2 &   4 &   2 & 0.5   & 0.809 & 0.958 & 0.919 & 0.958 \\
      2 &   5 &   2 & 0.4   & 1.039 & 1.321 & 1.175 & 1.321
  \end{tabular}
  \end{ruledtabular}
\end{table}

\begin{table}
  \caption{The parameter values $\lambda_s$, $\lambda_t$, and $\lambda_b$ 
    at the spinodal point and thermodynamic transition, 
    and at bifurcation of paramagnetic solution, respectively,  
    for $(K,C,L)$-MN codes over the BILC and various code parameters; 
    $\lambda_0$, denoting the information-theoretical upper bound 
    for error-free communication, is also shown.}
  \label{tab:MN-BILC}
  \begin{ruledtabular}
  \begin{tabular}{rrrddddd}
    $K$ & $C$ & $L$ & \mbox{$R$} 
    & \mbox{$\lambda_s$} & \mbox{$\lambda_t$} 
    & \mbox{$\lambda_b$} & \mbox{$\lambda_0$} \\\hline
      1 &   2 &   3 & 0.5   & 0.652 & 0.714 & - & 0.752 \\
      1 &   2 &   4 & 0.5   & 0.619 & 0.740 & - & 0.752 \\
      1 &   2 &   5 & 0.5   & 0.582 & 0.748 & - & 0.752 \\
      1 &   3 &   2 & 0.333 & 0.903 & 0.934 & - & 1.055 \\
      1 &   3 &   3 & 0.333 & 0.831 & 1.040 & - & 1.055 \\
      1 &   3 &   4 & 0.333 & 0.735 & 1.051 & - & 1.055 \\
      \noalign{\vspace{4pt}}
      2 &   3 &   2 & 0.667 & 0.525 & 0.551 & 0.597 & 0.553 \\
      2 &   3 &   3 & 0.667 & 0.464 & 0.553 & 0.493 & 0.553 \\
      2 &   3 &   4 & 0.667 & 0.419 & 0.553 & 0.437 & 0.553 \\
      2 &   4 &   2 & 0.5   & 0.689 & 0.751 & 0.771 & 0.752 \\
      2 &   5 &   2 & 0.4   & 0.807 & 0.914 & 0.894 & 0.914
  \end{tabular}
  \end{ruledtabular}
\end{table}

\begin{figure}
  \centering
  \begin{minipage}{75mm}
    \centering
    \includegraphics[width=75mm]{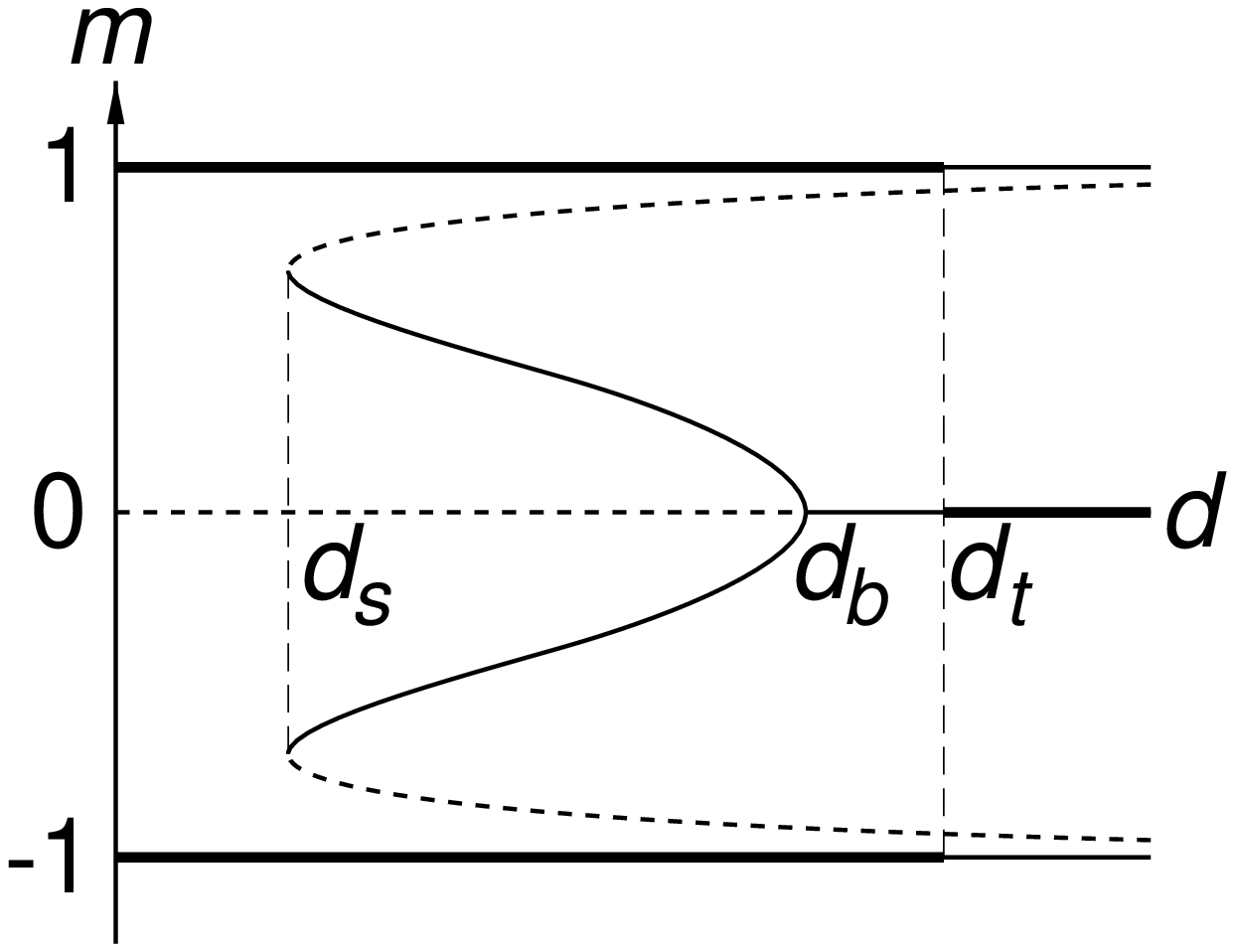}\\
    (a)
  \end{minipage}
  \quad
  \begin{minipage}{75mm}
    \centering
    \includegraphics[width=75mm]{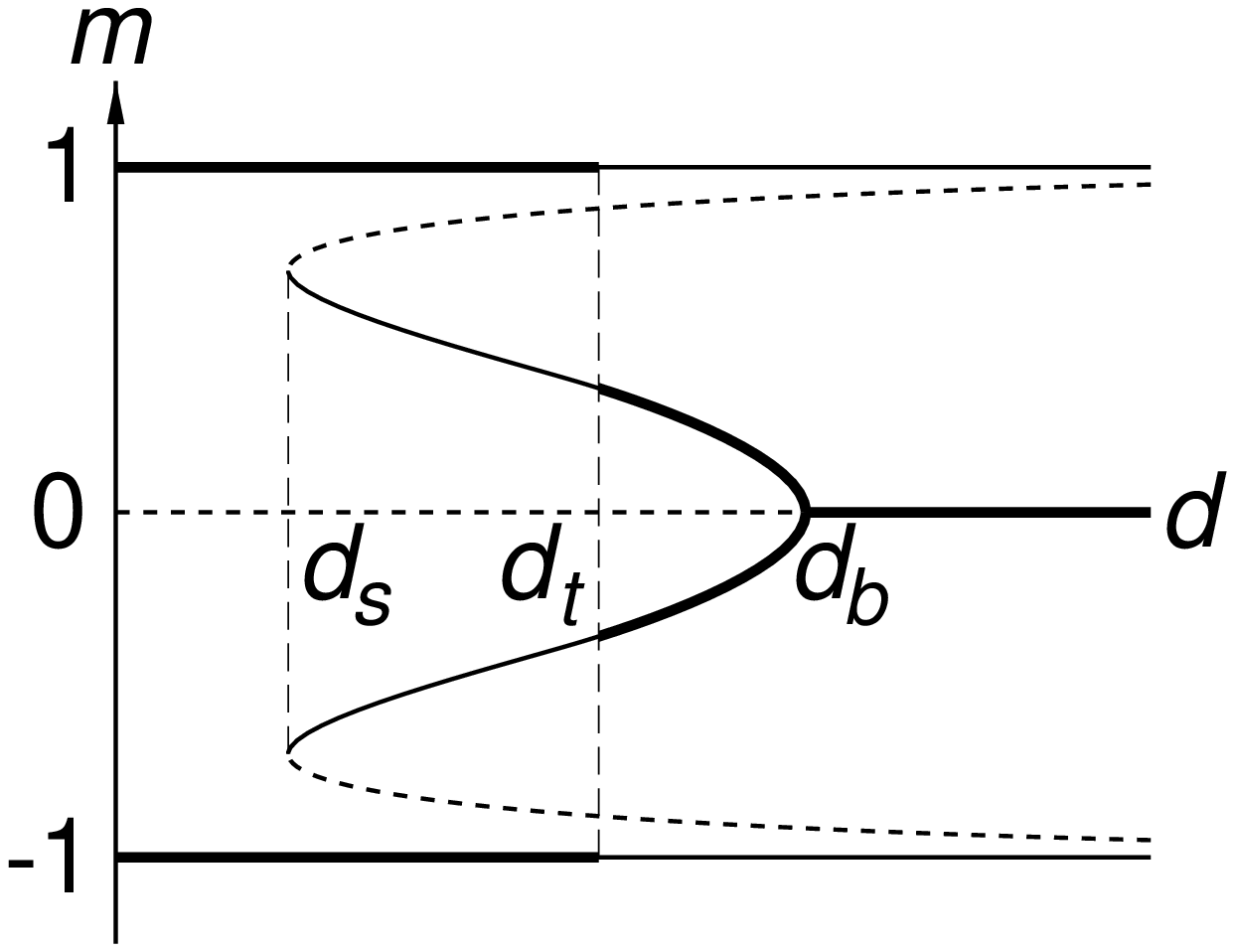}\\
    (b)
  \end{minipage}
  \caption{Noise-overlap diagram for the cases with $K=2$. }
  \label{fig:bif-diag-K2}
\end{figure}

The noise-overlap diagram for the cases with $K=2$ has the general
structure shown in Fig.~\ref{fig:bif-diag-K2}.  The diagram is
characterized by three transition points: the spinodal point 
$d_s$, the thermodynamic transition $d_t$, and the bifurcation point
$d_b$.  The order of the thermodynamic transition $d_t$ and the
bifurcation point $d_b$ varies with the values of $C$ and $L$, so that
the bifurcation pattern for the cases with $K=2$ is further divided
into two sub-categories depending on the order of the two transitions:
$d_s<d_b<d_t$ for the first group, and $d_s<d_t<d_b$ for the second
group.  The noise-overlap diagrams for these groups are illustrated in
Fig.~\ref{fig:bif-diag-K2} (a) and (b), respectively.  By the local
stability analysis the bifurcation point $d_b$ is determined by
\begin{equation}
  \label{eq:para-stability}
  \int v\rho(v)\,dv=(C-1)^{-1/L},
\end{equation}
which allows us to decide the type of bifurcation of a particular
case.  See the appendix for derivation of
Eq.~(\ref{eq:para-stability}).  As a result, we found that only a few
cases with small values of $C$ and $L$ fall into the second category.
The values of $C$ and $L$ for which the $(2,C,L)$-MN code fall into
the second category depend on the channel characteristics; as far as
we have observed, only the cases with $L=2$ fall into the second
group.  For the BIAWGNC case, the $(2,3,2)$-MN code is the only one
instance, whereas for the BILC case, both $(2,3,2)$- and $(2,4,2)$-MN
codes belong to this group.  (For the BSC case, $(2,3,2)$-,
$(2,4,2)$-, and $(2,5,2)$-MN codes belong to this group.)  All the
$(2,C,L)$-MN codes but those mentioned above are in the first group.
For the cases in the second group, 
the thermodynamic transition $d_t$ must be less than
the information-theoretic limit $d_0$: However, it turns out
numerically that $d_t$ is very close to $d_0$.

\begin{figure}
  \centering
  \includegraphics[width=75mm]{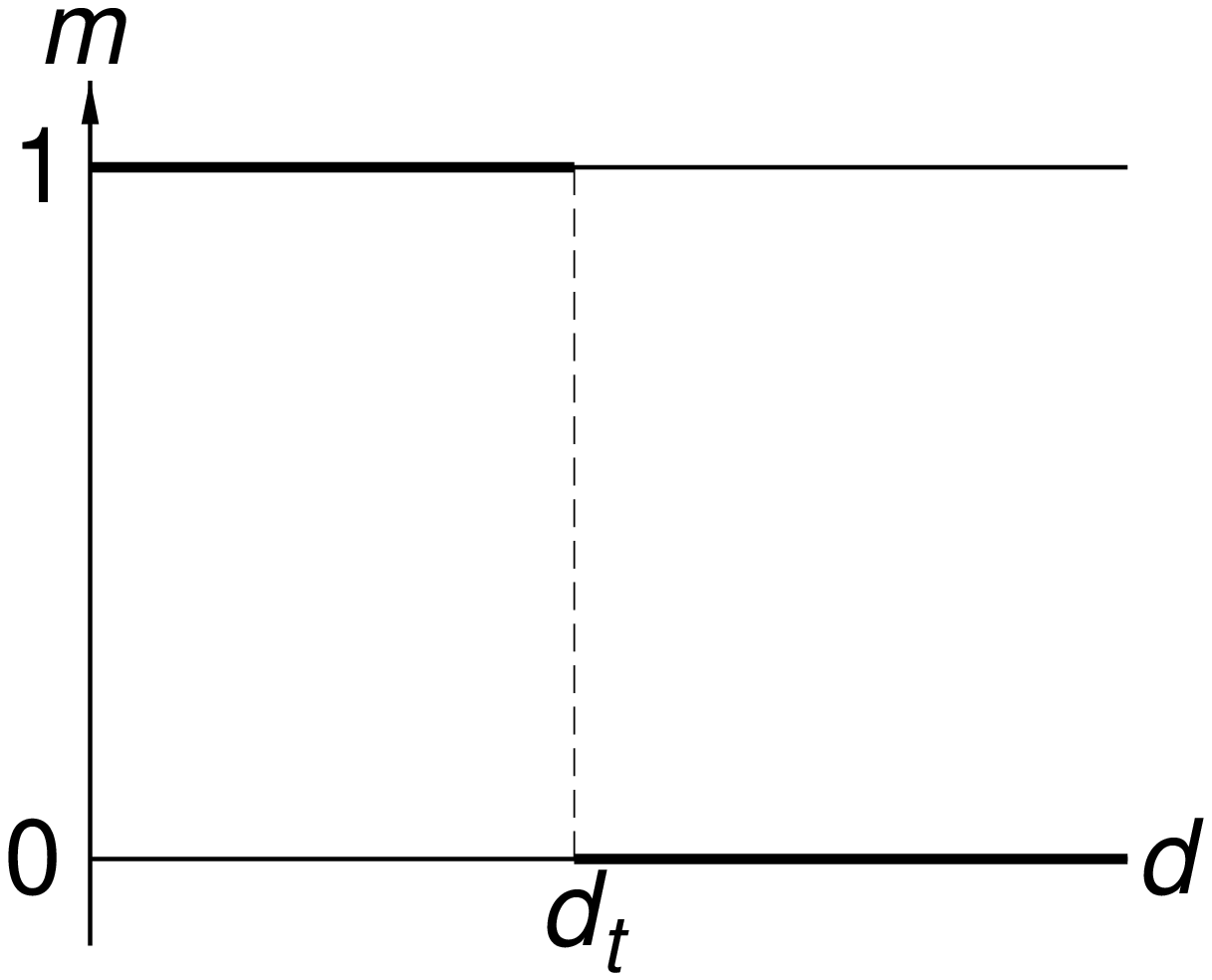}
  \caption{Noise-overlap diagram for the cases with $K\ge3$.}
  \label{fig:bif-diag-K3}
\end{figure}

We observed that the noise-overlap diagram for the cases with $K\ge 3$
is relatively simple for the BIAWGNC and BILC cases, just as in the
BSC case (Fig.~\ref{fig:bif-diag-K3}): The ferromagnetic solution with
$m=1$ (and its mirror image with $m=-1$ when $K$ is even) and the
paramagnetic solution are the only stable solutions found, both of
which are locally stable over the whole range of the noise level.  The
system exhibits a first-order transition at the information-theoretic
limit $d_t$.  We did not find any solutions other than the
ferromagnetic and paramagnetic solutions.

\section{Conclusions}
\label{sec:conclusion}

We have analyzed typical performance of LDPC codes 
over BIOS channel using statistical mechanics. 
We have shown for the case of LDPC codes 
that the log-likelihood ratio of the received signal 
serves as an external random field acting on each site, 
and that channel characteristics 
define the distribution of the random field. 
The Gallager and MN codes are analyzed, 
to find that the basic properties of these codes remain unchanged 
regardless of channel characteristics. 
In particular, it has been shown that 
these codes potentially saturate Shannon's limit 
asymptotically, as $K\to\infty$, for the Gallager code; 
and when $K,L\ge2$\ ---\ with a few exceptions with small $C$ and $L$ 
values --- and asymptotically as $L\to\infty$ for $K=1$, 
for the MN code.
Saddle-point solutions have also been numerically evaluated
extensively for the cases of BIAWGNC and BILC channels, from which
noise-overlap diagrams, as well as the transition and bifurcation
points, have been characterized.

\begin{acknowledgments}
  We would like to thank Yoshiyuki Kabashima for his helpful
  suggestions, Jort van Mourik for providing computer programs, and
  Nikos Skantzos for helpful discussions. Support from EPSRC research
  grant GR/N00562 is acknowledged.
\end{acknowledgments}

\appendix* 

\section{Stability of paramagnetic solution for $K\ge2$}

To probe the stability of paramagnetic solution, which exists for
$K\ge2$, we analyze the stability with respect to $\qq$ and $\rr$
only, and do not consider stability with respect to $\hat\qq$ and
$\hat\rr$; these conjugate variables are subsidiary to their
counterparts, $\qq$ and $\rr$, respectively, so that the former should
not be considered as independent variables.

Let $A$, $B$, $\ldots$ denote sets of replica indices such as
$\langle\alpha_1\cdots\alpha_m\rangle$, $m\ge1$.  We first evaluate
the Hessian of the free energy (\ref{eq:MN-free-energy}) with respect
to $4\times(2^n-1)$ variables $\{q_A,\hat q_A,r_A,\hat r_A\}$:
\begin{equation}
  H=\left(
    \begin{array}{cc}
      \begin{array}{cc}
        H_{\qq \qq} & H_{\qq \hat\qq} \\
        H_{\qq \hat\qq} & H_{\hat\qq \hat\qq} 
      \end{array}
      & \hbox{\Large$O$} \\
      \hbox{\Large$O$} & 
      \begin{array}{cc}
        O & H_{\rr \hat\rr} \\
        H_{\rr \hat\rr} & H_{\hat\rr \hat\rr}
      \end{array}
    \end{array}\right),
\end{equation}
where 
\begin{eqnarray}
  \bigl(H_{\qq \qq}\bigr)_{AB}
  &=&\left\{
    \begin{array}{ll}
      0 & (K\ge3)\\
      \displaystyle
      -\frac{C}{q_0^2}\left(\frac{r_A}{r_0}\right)^L\delta_{AB} & (K=2)
    \end{array}\right.
  \nonumber\\
  \bigl(H_{\qq\hat\qq}\bigr)_{AB}
  &=&\delta_{AB}
  \nonumber\\
  \bigl(H_{\hat\qq\hat\qq}\bigr)_{AB}
  &=&-\frac{C(C-1)}{\hat q_0^2}\delta_{AB}
  \nonumber\\
  \bigl(H_{\rr\hat\rr}\bigr)_{AB}&=&\frac{M}{N}\delta_{AB}
  \nonumber\\
  \bigl(H_{\hat\rr\hat\rr}\bigr)_{AB}
  &=&-\frac{M}{N}\frac{L(L-1)}{\hat r_0^2}\delta_{AB}
\end{eqnarray}
The block-diagonal structure of the Hessian 
allows us to decompose the stability problem 
into two, one with respect to $\qq$, 
and another with respect to $\rr$. 

Following the argument in the appendix of~\cite{Tanaka-ITpre}, one can say
that the system is stable with respect to $\qq$ if the matrix
$H_c\equiv H_{\qq\qq}
-H_{\qq\hat\qq}\bigl(H_{\hat\qq\hat\qq}\bigr)^{-1}H_{\qq\hat\qq}$ is
positive definite.  This condition takes into account the fact that
$\hat\qq$ depends on $\qq$.  A corresponding statement holds for the
stability with respect to $\rr$.

The stability with respect to $\rr$ is straightforward, by noting that
the matrix $H_{\hat\rr\hat\rr}$ is negative definite, which means that
$H_c=-(M/N)^2\bigl(H_{\hat\rr\hat\rr}\bigr)^{-1}$ is positive
definite.

We consider the stability with respect to $\qq$.  For $K\ge3$, we have
$H_c=[\hat q_0^2/C(C-1)]I$, where $I$ is the identity matrix, so that
the stability immediately follows, irrespective of the noise level of
the channel.  For $K=2$, the matrix $H_c$ is diagonal, and its $A$-th
element is
\begin{equation}
  \bigl(H_c\bigr)_{AA}=-\frac{C}{q_0^2}
  \left(\frac{r_A}{r_0}\right)^L+\frac{\hat q_0^2}{C(C-1)}.
\end{equation}
Using the equality which holds under the RS ansatz, 
\begin{equation}
  \frac{r_A}{r_0}
  =\int_{-1}^1 v^m\rho(v)\,dv,
\end{equation}
where $A=\alpha_1\cdots\alpha_m$, 
we have, as the stability condition, 
\begin{equation}
  E_m\equiv\int_{-1}^1 v^m\rho(v)\,dv<(C-1)^{-1/L}.
\end{equation}
for $m=1,\ldots,n$. 
Since it can be shown that $E_{2m-1}=E_{2m}$ 
and $E_{2m}\ge E_{2m+2}$, 
the critical condition determining the stability is 
\begin{equation}
  E_1<(C-1)^{-1/L}.
\end{equation}

\end{document}